# Accelerating Variational Quantum Algorithms Using Circuit Concurrency

Salonik Resch[♦][1]
resc0059@umn.edu

Anthony Gutierrez
anthony.gutierrez@amd.com

Joon Suk Huh[‡]
jhuh23@wisc.edu

Srikant Bharadwaj
srikant.bharadwaj@amd.com

Yasuko Eckert
yasuko.eckert@amd.com

Gabriel Loh
gabriel.loh@amd.com

Mark Oskin
mark.oskin@amd.com

Swamit Tannu[‡]
stannu@wisc.edu

[♰]AMD Research

[♦]University of Minnesota-Twin Cities

[‡]University of Wisconsin-Madison

## Abstract

*Variational quantum algorithms (VQAs) provide a promising approach to achieve quantum advantage in the noisy intermediate-scale quantum era. In this era, quantum computers experience high error rates and quantum error detection and correction is not feasible. VQAs can utilize noisy qubits in tandem with classical optimization algorithms to solve hard problems. However, VQAs are still slow relative to their classical counterparts. Hence, improving the performance of VQAs will be necessary to make them competitive. While VQAs are expected perform better as the problem sizes increase, increasing their performance will make them a viable option sooner. In this work we show that circuit-level concurrency provides a means to increase the performance of variational quantum algorithms on noisy quantum computers. This involves mapping multiple instances of the same circuit (program) onto the quantum computer at the same time, which allows multiple samples in a variational quantum algorithm to be gathered in parallel for each training iteration. We demonstrate that this technique provides a linear increase in training speed when increasing the number of concurrently running quantum circuits. Furthermore, even with pessimistic error rates concurrent quantum circuit sampling can speed up the quantum approximate optimization algorithm by up to 20× with low mapping and run time overhead.*

## 1. Introduction

The era of quantum computing is arriving with rapid and consistent improvements in qubit technology. Quantum computers, or quantum processing units (QPU), with tens of qubits already exist and those with hundreds and even thousands are only a few years away [14]. Despite the advances in qubit technology, we will remain in the noisy intermediate-scale quantum (NISQ) [27] era for years, where noisy qubits limit circuit depth and error correction is not feasible. Running without error correction increases the speed of progress but places strong limitations on the number of qubits and depth of the quantum circuits that can be run [27][37]. We use the terms "circuit" and "program" interchangeably throughout this paper.

Variational quantum algorithms (VQA) [6] are viewed as a viable approach to achieve quantum advantage [12][26] on real problems in the NISQ era because of their hybrid nature. VQAs use parameterized quantum circuits and rely on classical processors for parameter optimization. Treating QPUs as accelerators in this manner allows for solving or approximating solutions for nearly all envisioned quantum computing problems with relatively shallow circuit depth, thereby making them resilient against noise.

However, VQAs are still slow relative to their classical algorithm competitors [15]. Hence, they are not yet state-of-the-art for solving problems of interest. Quantum operations are slow relative to classical computers, and the bottleneck for performance is the time it takes to collect samples from the quantum computer. In our benchmarks, even assuming the relatively fast operations on superconducting quantum computers [22], we find that the quantum sampling takes approximately 167× longer than classical processing and optimization. This is exacerbated for other quantum technologies, such as Ion-Trap and Neutral Atom, which have latencies that are orders of magnitude longer than superconductors [22]. These slow operations lead to long latencies for VQAs for problems of useful sizes [19]. Further, these long latencies can lead to degradation in the output quality. For long-running VQA algorithms, the physical characteristics of the qubits can change significantly during the length of the program (qubit drift). This introduces the need for periodic re-calibrations [19].

Additionally, due to limited coherence time and high gate errors, even the shallow depth VQAs cannot fully utilize all qubits in emerging QPUs with hundreds or thousands of qubits. Figure 1(a) shows the estimated success probability

---

[1] This work was completed while the author was an intern at AMD Research.

[24][34] for running the quantum approximate optimization algorithm (QAOA) [11] with qubit error rates from current IBM and Google QPUs (average gate fidelity = 0.99) [2][8] and optimistic projections of future QPUs (average gate fidelity = 0.999). Note that both IBM and Google QPUs currently have about 50 qubits, and we extrapolate them out to 1,000 qubits assuming the same noise rates. Under realistic noise assumptions there is effectively zero chance to operate a program with more than a few hundred qubits without accounting for error. Noise will degrade the quantum state over time and programs that are too large, and long-running will produce output that is too erroneous to be useful. Hence, even with large QPUs NISQ algorithms are limited to relatively small and short-running programs. Additionally, training large VQAs is challenging due to vanishing gradients with noisy qubits [37].

Hence, running a single program on a QPU in the NISQ era will have a high latency, be vulnerable to qubit drift, have a low probability of success, and will significantly underutilize expensive hardware as many of the qubits will remain idle. Circuit-level concurrency, running multiple quantum circuits simultaneously, is a natural solution to this problem. To that end, we propose *Concurrent Quantum Circuit Sampling (CQCS)*, which can increase the performance of quantum algorithms by making use of qubits that would otherwise be wasted. VQA applications involve repeated execution and measurement (sampling), where each repeated execution is entirely logically independent. This introduces the opportunity exploit parallelism by splitting a single quantum program into multiple instances. Running each instance in parallel can effectively increase the sampling rate. Hence, circuit-level concurrency can be leveraged to reduce the runtime of VQAs.

To be of use, the net impact of circuit concurrency must be positive (i.e., the error remains at a tolerable level). Another challenge is the scalability of the classical resources required. As quantum computers get larger, the process of finding optimal mappings for each program becomes a challenge for the supporting classical hardware. The qubit mapping process must be done prior to the execution of every program and is therefore latency sensitive. However, finding the best possible mapping is critical for the success of the quantum programs, hence significant classical computation should be invested to produce the best possible mapping. In the CQCS execution model the mapping process is embarrassingly parallel and can be easily scaled for applications using hundreds of qubits.

In this work, we demonstrate the effectiveness of circuit concurrency to increase the sampling rate for modern VQAs. Specifically, we leverage circuit concurrency to increase the training speed of QAOA. We test this strategy under multiple noise models and show that circuit concurrency can be beneficial even in the presence of realistic noise levels. The contributions of this work are as follows:

o We demonstrate that utilizing circuit-level concurrency can provide up to 20× speedup for QAOA.

o We show that circuit-level concurrency can be effective in the presence of noise.

## 2. Background
### 2.1. Quantum Computing

Quantum Computers are envisioned as co-processing units that speed up a class of applications by leveraging the properties of quantum bits (qubits). By using qubit registers (collection of qubits) and applying a sequence of quantum instructions (gates) on the qubit registers, programmers can run quantum algorithms. Typically, quantum registers are used to represent probability distributions. For example, $N$ qubits can represent a probability distribution with $2^N$ possibilities. A quantum circuit can transform the input problem distribution into an output distribution that encodes the problem's solution. However, this distribution is not directly accessible as reading $N$ qubits yields a single $N$-bit binary string with a certain probability. To estimate the output distribution, we repeatedly execute the circuit for many trials. Reading a qubit is identical to sampling from a probability distribution.

### 2.2. Errors on Quantum Computers

On a physical level, superconducting circuits, trapped ions, or semiconductor devices can be used as qubits. Unfortunately, all existing qubit technologies are highly susceptible to errors. On both IBM and Google superconducting quantum computers, the average gate error rate is about 1% and readout error is about 2% [2][8]. Furthermore, a qubit's state decays within hundreds of microseconds, limiting the number of computations we can perform coherently. Moreover, current quantum computers show a high variability in error rates.

Quantum hardware's reliability is steadily improving. Both device and software innovations have reduced errors by 5× since IBM Quantum Cloud Service was launched in 2016 [8]. However, running quantum algorithms with over 1,000 quantum operations is beyond the reach of current devices, which are limited by short qubit coherence time. To scale quantum applications, we will need a mechanism to extend qubit lifetimes by continuously detecting and correcting qubit errors. Quantum error correction (QEC) can enable such a fault-tolerant paradigm. It is based on redundancy computation by creating a fault-tolerant logical qubit that uses multiple physical qubits. Unfortunately, QEC is expensive. We need many thousands of physical qubits to enable fault-tolerant quantum applications. In the near-term, quantum computers with thousands of qubits will operate in the NISQ computing mode.

In the NISQ era, gate and decoherence errors corrupt the program state and produce incorrect outcomes. The circuit noise distorts the shape of an output probability distribution. For example, consider a bell pair generation circuit that is used to entangle two qubits. In theory, this circuit produces an output distribution $P_{ideal}$ = ["00" : 0.5; "11" : 0.5], where outcomes "00" and "11" are equally likely. However, when executed on NISQ hardware the circuit could produce an output distribution such as $P_{noisy}$ = ["00" : 0:55; "11" : 0:30; "10" : 0:10; "01" : 0:05]. Hardware errors corrupt the program and produce erroneous outcomes ("10" and "01") to skew the output distribution. With increasing error rate, the difference (distance) between $P_{ideal}$ and $P_{noisy}$ increases. With no error correction, errors accumulate on NISQ hardware as the number of gates increases.

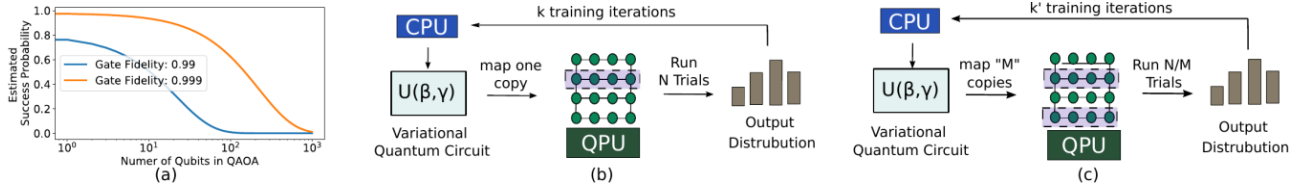

**Figure 1: (a) Estimated success probability of QAOA for max-cut with 3-regular graph as input (b) Baseline approach of running VQA (c) Concurrent execution of M copies of VQA to improve sampling rate.**

### 2.3. Variational Quantum Algorithms

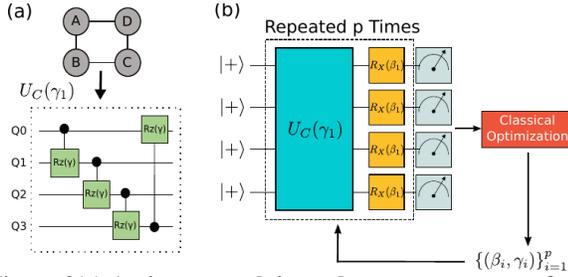

**Figure 2(a) An input graph is used to construct part of the parametric circuit in QAOA (b) QAOA uses a parametric circuit block repeated p times to generate output.**

QAOA is one such variational algorithm that can be executed on noisy QPUs. QAOA can solve max-cut, max-clique, and 3-SAT problems, well-known NP-hard combinatorial optimization problems. This paper focuses on QAOA applied on max-cut, which is a promising candidate to show quantum advantage. We consider undirected 2- and 3-regular graphs, standard practice in the literature [40].

The input can be formulated as a graph, as shown in Figure 2(a), where vertices represent variables and edges between vertices represent constraints [28]. The number of vertices (variables) in the input graph corresponds to the number of qubits in the corresponding quantum circuit. Effectively each variable is assigned to a qubit. Edges between vertices specify 2-qubit controlled rotation gates, which are to be performed on the corresponding qubits.

The QAOA circuit, shown Figure 2(b), begins by applying a Hadamard (H) gate on all qubits. Controlled rotations are then performed for each edge of the input graph, implemented by two CNOT gates and single-qubit z-rotations (Rz). The angle of rotation is specified by the classical parameter $\gamma$. Then, single qubit x-rotations (Rx) are performed on all qubits, with the angle specified by the classical parameter $\beta$. The sequence of controlled z-rotations and x-rotations is then repeated a specified number of times.

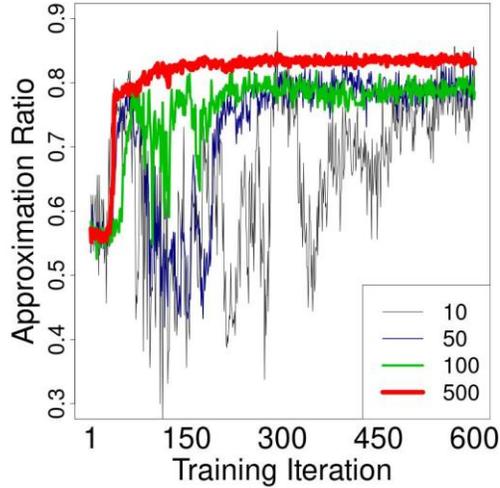

**Figure 3: Impact of the number of samples per training iteration on training time and quality of solution for QAOA with no quantum noise.**

SWAP gates, which are gates that swap state between qubits, may also be necessary if the QPU topology does not provide all-to-all connections. The classical hyperparameter $p$ determines the number of repetitions. Each stage has a separate $\beta$ and $\gamma$. Circuits with larger $p$ can produce better results, however they are deeper and therefore more susceptible to noise. For a given noise rate there will be an optimal value of $p$ [37].

Measurement at the end of the circuit provides a bit string which is a solution to the input problem. The quality of this solution (score) can be computed efficiently by a classical computer. The QAOA circuit will be run many times with a bit string being produced on each run. The average score of the bit strings produced is called the expectation value. If the best solution to the input problem is known, which is true for small problems, we can compute the *approximation ratio*, which is the expectation value divided by the best solution. The approximation ratio is used as a metric to determine the quality of a given QAOA circuit. If the approximation ratio is 1.0, the circuit produces a bit string with the best answer 100% of the time. The approximation ratio is guaranteed to be greater than 0.6924 for a QAOA circuit with $p = 1$ and increases monotonically with increasing circuit depth [11].

The optimal classical parameters $\beta$ and $\gamma$ are unknown initially. Hence, they must be learned over time. This can be done by rerunning the circuit many times and using a classical optimizer to update the parameters. Hence, as we learn the parameters over time, we expect to see an increase in the approximation ratio. The final approximation ratio will plateau at some point having reached its best possible solution or because noise limits further improvement.

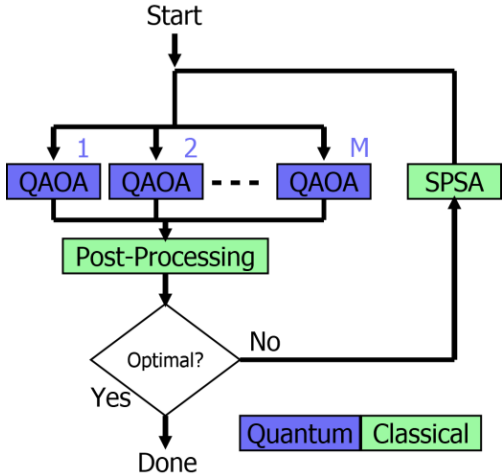

**Figure 4: CQCS uses M concurrent instances of QAOA to linearly increase the number of samples collected per unit time.**

In the near term, VQA can enable resource-efficient quantum circuits to solve practical problems [15]. However, hardware errors pose a significant challenge in obtaining reasonable quality solutions. How to enable high-quality solutions using VQA on NISQ machines is an open problem, and many prior works focus on addressing this challenge by improving device reliability, algorithm design, and compilation strategies [1][21][34][40][41]. With innovations across the computing stack, it is possible to reach the desired quality of solution with NISQ machines [19]. However, to enable quantum speedup, we must consider both the quality of the solution and the total execution time of VQA. In this paper, we focus on the latter.

## 3. Challenges with Accelerating VQA

In this section, we will discuss key challenges in enabling quantum speedup for VQAs on near-term QPUs.

### 3.1. VQAs are Latency Sensitive

Shor's algorithm [31] for integer factorization promises exponential speedup by reducing computational complexity. It guarantees the exact solutions in the absence of qubit noise. In contrast, VQAs provide approximate solutions that may not be optimal but are nevertheless high-quality. VQAs can be treated as a heuristic solution like classical approximate algorithms. For many discrete optimization and quantum chemistry problems, VQAs are believed to outperform classical heuristics that are currently used by practitioners. For an end-user, a VQA is an attractive proposal if the VQA can reduce average time to solution. Thus, to show the quantum advantage, VQAs must outperform classical heuristics. In this paper, we focus on reducing the latency of

VQAs by introducing the notion of circuit-level parallelism. In the subsequent sections, we will discuss bottlenecks in VQAs and how they scale with input size.

### 3.2. Modeling Execution Time of VQAs

As shown in Figure 1(b), VQAs are executed in three steps. **(1)** Execute a quantum circuit for a set of parameter values ($\beta$ and $\gamma$ in the case of QAOA) $N$ times to collect $N$ samples. **(2)** Compute the gradient in expected value of the cost function using $N$ samples to find new $\beta$ and $\gamma$ values. (3) Run the optimization loop for $K$ iterations to converge on the desired $\beta_{opt}$ and $\gamma_{opt}$ values. Using these steps, a VQA's execution time can be modeled using the number of iterations ($K$), optimizer delay ($T_{opt}$), number of samples ($N$), and time to generate one sample ($T_s$).

$$Execution\ Time = K \times T_{opt} + N \times T_s$$

Parameters $K$ and $T_{opt}$ depend on the optimization algorithm and the optimization problem landscape. Whereas the spread (variance) of output probability distribution determines $N$. $T_s$ is the time required to generate a sample by running a quantum circuit once. It can be modeled as the sum of circuit latency ($T_{circ}$) and circuit processing delay ($T_{proc}$).

$$T_s = T_{circ} + T_{proc}$$

$T_{circ}$ is proportional to circuit depth and average gate latency. In contrast, $T_{proc}$ is the delay introduced by pre-processing and post-processing steps. Typically, $T_{proc}$ does not depend on the size of the program and can be amortized by a pipelined design. For circuits with tens of qubits, $T_{proc} \ll T_{Circ}$.

$$T_s \approx T_{circ} = O\big(poly(Num\ qubits)\big)$$

Learning the optimal parameters relies on having good estimates of the approximation ratio during the training process. Enough samples from the output of the quantum circuit must be taken to achieve this. As we are limited to some finite number of samples, sampling error, in addition to quantum noise, plays a significant role in the training process [33]. We show the training process of QAOA for different sampling rates in Figure 3. As can be seen, taking too few samples result in unstable behavior. When solving relatively small problems, such as those which can be solved by existing QPUs, a high level of instability can be tolerated. The training process is still likely to find a near-optimal solution. This instability is undesirable as larger and deeper circuits (which require higher precision) will be unable to tolerate the significant output noise.

As simulations with too few samples produce erratic and unpredictable results, the rest of the paper includes only simulations that have well above the required number of samples. However, it should be noted that the number of samples per training iteration is a variable parameter that can be tuned to a specific problem. The number of samples required heavily depends on the number of variables, the noise rate, and optimization hyperparameters, such as the learning rate.

## 4. Insight and Design

We propose CQCS, to leverage concurrency to increase the number of samples collected per unit time for VQAs.

### 4.1. Opportunities for Concurrent Execution

As discussed in Section 2.3, VQAs typically require many iterations and measurements of the quantum circuit to produce a usable result. Enough samples must be gathered to reliably solve for more optimal parameters. Collecting these samples from the QPU is the most significant contributor to the total latency, the classical optimization only accounts for a small fraction. For example, collecting a single sample from a 12-qubit quantum circuit implementing QAOA takes roughly 25μs (based on reported IBM gate latency and the circuit depths used in the work), which must be repeated typically a few thousand times. If we assume 1,000 samples, this will require 25ms. In contrast, classical optimization requires only a few arithmetic operations per sample. Processing all 1,000 samples takes less than 150ms. Hence, sampling the quantum circuit takes over 99% of the total latency. This time difference will be even more pronounced with alternative quantum technologies. For example, Ion-Trap computers have gate latencies that are 150-800× longer than superconductors [13][22][30]. This means the performance of VQAs is limited by the time overhead of sampling.

The generation of each sample is entirely independent, introducing the opportunity for parallelization. To execute the generation of each sample in parallel, the VQA circuit must be split into different instances, with each running on a different set of physical qubits. Splitting into $M$ instances can increase the sampling rate up to $M$ times, which has the potential to reduce the latency by up to a factor of $M$. The number of instances that can be in run in parallel will be limited by the hardware available; there is no algorithmic limitation. In traditional terms, there are no *data hazards*, only *structural hazards*. **The lack of data hazards allows us to exploit higher degrees of parallelism.** However, structural hazards for QPUs are more complicated than for classical computers. We not only need to know if the resources (qubits) are available, but also if they are of sufficient quality. **Hence, as we increase parallelism, we must account for potential losses in quality due to quantum structural hazards.**

### 4.2. Leveraging Underutilized Qubits

**The race for building larger QPUs:** Currently, many companies and university labs are racing towards building larger QPUs. The primary goal of current efforts is to demonstrate fault-tolerant logical qubits. To that end, companies are aiming to build quantum computers with thousands of physical qubits. Recently, IBM has announced their roadmap to scale quantum hardware to over 1,000 qubits by 2023 [7]. However, even with thousands of qubits, quantum algorithms cannot operate in a fault-tolerant regime to solve practical problems. For that, we will need tens of thousands of qubits [29].

**Limit on scaling NISQ Applications:** High gate error rates, limited coherence time, and large SWAP overhead due to limited connectivity are three bottlenecks that limit the size of NISQ applications. Although many NISQ applications like VQA require shallow circuits, they still rely on creating large entangled states. Unfortunately, entangling qubits on a NISQ machine is expensive due to limited connectivity. To entangle multiple qubits on quantum hardware, SWAP gates are inserted. Unfortunately, SWAP gates increase the error and significantly reduce the odds of success. Even for resource-efficient VQAs, the number of SWAP gates increases quickly with the circuit size [16]. In the near term, our analysis shows that quantum applications will be limited to at most 100 qubits (illustrated in Figure 1(a)). Even when we have 1,000 qubit quantum computers, only a fraction of the qubits can be utilized for running NISQ applications.

### 4.3. CQCS Design

As an example, if a VQA requires $N$ samples of a quantum circuit, the $N$ samples can be split across $M$ different instances. Enough samples can be taken if each individual instance is sampled $\left\lceil \frac{N}{M} \right\rceil$ times. If $M$ divides $N$, the samples from each instance are of equal quality and each circuit is of equal length, this provides a speedup of $M$. However, as will be discussed in Section 4.4, noise can cause each instance to be of unequal quality, which could result in subpar accuracy of the distribution and slow the training process. Additionally, as will be discussed in Section 4.5, each instance may be of unequal lengths as well. This can increase the latency of sampling. However, we will show in Section 6 that these drawbacks do not remove the benefit of CQCS.

### 4.4. Impact of Noise on CQCS

Modern QPUs have high noise rates, and the rates on different qubits on the same QPU can vary considerably. Additionally, the noise rate for single-qubit gates, 2-qubit gates, and measurements can all be different for the same qubit. This variability makes the mapping from logical qubits to physical qubits critical for the success of a quantum circuit. The mapping process has a complex trade-off space, even for a single circuit, and the complexity naturally increases for multiple circuits. An unfortunate, and unavoidable, consequence is that the average noise rate of the utilized qubits will be higher for CQCS. This is because there is a finite number of high-quality qubits on the computer, which must be shared amongst simultaneously active circuits. Hence, there is a trade-off between the amount of circuit concurrency and the noise rate per qubit.

Coherent and correlated noise poses a particularly strong threat to CQCS. It is known that coherent noise is dominant

| Noise Model | Single Qubit Noise | Two Qubit Noise | Measurement Error |
|---|---|---|---|
| Pauli (P) | 0.06 – 0.24% | 0.07 – 0.65% | 1.16 – 4.95% |
| Amplitude/Phase Damping (APD) | 0.06 – 0.24% | 0.07 – 0.65% | 1.16 – 4.95% |
| Coherent (C) | $0.025\pi - 0.05\pi$ | $0.025\pi - 0.05\pi$ | 1.16 – 4.95% |

**Table 1: Parameters of noise models used in simulation. Pauli and APD noise and measurement error rates are based on Google's 52-qubit Sycamore machine [2], matching the mean and standard deviation. Noise rates listed are used for all QPU architectures.**

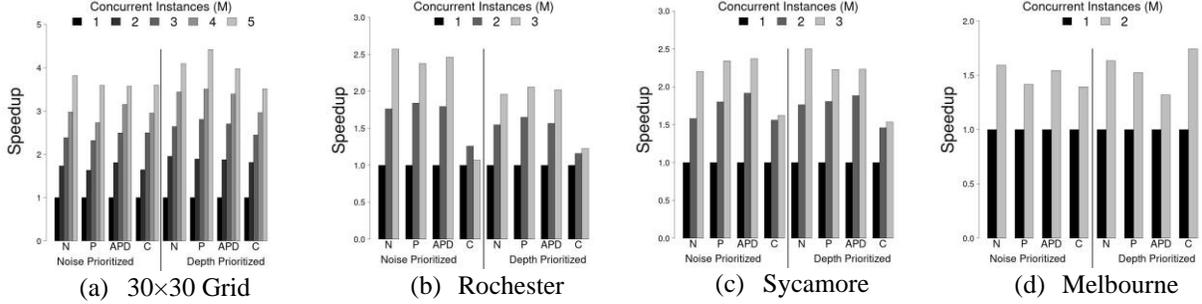

(a) 30×30 Grid    (b) Rochester    (c) Sycamore    (d) Melbourne

**Figure 5: Speedup with M concurrent instances on different architectures, with Noiseless (N), Pauli (P), Amplitude and Phase Damping (APD), and Coherent (C) noise. Results are shown for both noise prioritized and depth prioritized scheduling.**

[5] and has worst-case impact on the quantum state [36]. However, it can be especially damaging for CQCS. Consider Pauli noise, which represents a stochastic noise process. If multiple instances of a program each experience varying levels of Pauli noise, there will be varying levels of noise on their output distributions. However, these outputs still have the same underlying distribution. If multiple instances experience different forms of coherent and correlated noise, the outputs can be effectively different distributions. This makes it more difficult to combine their outputs.

Using more qubits can increase the risk of crosstalk noise. While this is a problem, the multiple, smaller instances of a program will be less susceptible than one large program. This is because these smaller instances will require smaller entangled states, have fewer logical operations, and have lower coherence time requirements. Regardless, the noise rate will increase with simultaneously executed operations. Experimental data from Google's 52-qubit Sycamore processor provides a good case study for this effect [2]. Going from performing gate operations in isolation to full utilization increases the single qubit gate noise by 6.67%, the 2-qubit gate noise by 43.1%, and measurement errors by 23.6%.

### 4.5. Mapping in CQCS

As described in Section 2.1, mapping logical quantum circuits to physical qubits typically adds SWAP gate overhead. This SWAP overhead applies to single circuits but adds some additional complexity for CQCS. The number of SWAPs required significantly depends on the connectivity of the physical qubits. As different instances are mapped to different physical qubits, each instance can have a different SWAP overhead. This variance in SWAP overhead will increase the length of some instances more than the others. A direct consequence of this is that the classical processor will have to wait longer for results from some instances than the others. As the classical processor must receive results from all instances before performing optimization, it will have to wait for the slowest instance. We refer to the latency of the slowest instance as $T_s'$, which sets the effective sampling latency. Hence, the latency of a training iteration will be increased by $\frac{T_s'}{T_s}$ due to SWAP overheads.

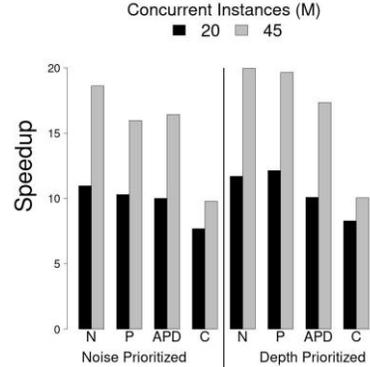

**Figure 6: Speedup with high levels of concurrency, $M = 20$ and $M = 45$, on the 30×30 Grid architecture.**

We refer to the $\frac{T_s'}{T_s}$ ratio as the *sample slowdown*. The sample slowdown will be significantly impacted by the connectivity of the QPU architecture. On QPUs with high connectivity, different instances of a program are more likely to have to similar SWAP overheads. In architectures with sparse connectivity, which tend to have only a few regions of higher connectivity, each instance is more likely to run on a significantly different topology, and consequently have more widely varying swap overheads.

The change of effective sample latency from $T_s$ to $T_s'$ adversely impacts the benefit of concurrency by increasing the iteration latency. This counteracts the reduction in the iteration latency that comes from running multiple simultaneous instances. By running $M$ instances of QAOA concurrently, the latency to generate the total number of samples can be reduced by *M* times, but each sampling latency is also increased by $\frac{T'_s}{T_s}$. The net result is:

$$Relative\ iteration\ latency = \frac{1}{M} \times \frac{T'_s}{T_s}$$

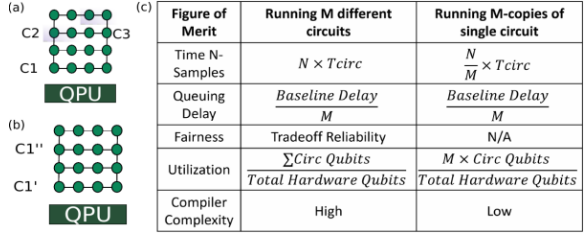

**Figure 7: (a) Multiprogramming model (b) CQCS model (c) Comparison.**

### 4.6. CQCS vs Multiprogramming

Previous work has considered multiprogramming (MP) on quantum computers [10][23] with the goal of increasing qubit utilization. This involves running different quantum programs, potentially from different users, at the same time. While this approach can work, it comes with some unique challenges. The most complex challenge is to ensure fairness by providing each user equal service. As the quality of physical qubits on the QPU can vary, there is an unavoidable competition between different programs for more reliable resources. Hence, any MP scheduler should carefully analyze this trade-off space and will run the risk of providing subpar service to one or more users. This requires the scheduler to monitor the output quality of each program and throttle the amount of parallelism if it detects significant degradation [10].

Additionally, running multiple programs simultaneously will create on overall higher level of noise due to crosstalk, which can vary depending on the types of operations that are performed at the same time. Hence, one user's program may be directly adversely affected by that of another, and the user will have no knowledge as to what other programs are currently executing. Another challenge for MP is that different programs can be of significantly different lengths. As measurements need to be performed at the same time, programs need to be scheduled to complete at the same time [10]. This is analogous to the increase in sampling latency, Ts, discussion in Section 4.5. This reduces the amount of exploitable parallelism, as shorter programs will be delayed for longer ones.

CQCS avoids much of this complexity while still increasing qubit utilization. CQCS runs multiple instances of a single program simultaneously. Since all instances are from the same program, and hence the same user, there is no fairness concern. A single user will maintain exclusive access to the QPU at any given time. This reduces the complexity of the scheduler. Running more instances will increase the level of noise due to crosstalk, just as with more programs in MP.

However, a user can control how many simultaneous instances run and can tune the configuration to meet their needs. Additionally, as all instances are from the same program, they will have similar circuit depths. This helps maintain a minimal sampling latency, $T_s$. Comparison of MP and CQCS is shown in Figure 7.

## 5. Methodology

We assume the system consists of a classical host processor and a QPU. All code executed on the host processor is written in C++. We use the Qiskit simulator [9] to emulate the QPU and the host-side C++ code interfaces with Qiskit over pybind11 [18]. To model noise, we use a variety of models, including Pauli, Amplitude/Phase Damping [35], and coherent rotations [5]. We use a variety of QPU architectures, including IBM's 16-qubit Melbourne, 53-qubit Rochester, Google's 53-qubit Sycamore, and hypothetical 900-qubit 30x30 grid.

We use 12-qubit QAOA circuit for the Grid, Rochester, and Sycamore architectures and a 4-qubit QAOA circuit for the Melbourne architecture. For each size we use a single, randomly generated input graph, the 4-qubit graph is 2-regular and the 12-qubit graph is 3-regular. Literature on QAOA typically uses 3-regular graphs, and the use of a single input graph is acceptable due to a concentration of the optimization landscape for different random graphs [5][33].

We use random initialization of the parameters $\beta$ and $\gamma$, and for optimization we use the simultaneous perturbation stochastic approximation (SPSA) algorithm [32]. SPSA has been shown to have robust and fast convergence relative to other optimization algorithms for QAOA in the presence of quantum noise [33]. The host processor performs the classical optimization between calls to the QPU.

The host processor launches all the instances as jobs onto the QPU, which run in parallel, and waits for the measurement results. Many samples (repetitions of the circuit) will be collected for each instance before returning the results to the host. The host performs post-processing, combines the results from each instance, and uses the SPSA algorithm to update the parameters using gradient descent. This process is repeated many times. Under practical circumstances, the terminating condition is when a sufficiently optimal solution is found. However, with our small QAOA problems (which can be efficiently simulated) optimal solutions are measured frequently. Hence, we continue the training process until the maximal approximation ratio is reached.

### 5.1. Noise Modeling

We use three types of noise models: Pauli (P), amplitude and phase damping (APD), and coherent (C). Both Pauli and APD are standard noise models, for which we use the built-in Qiskit implementations. The Pauli noise model involves inserting X, Y, and Z gates into the circuit at random. This is like a classical bit flip model but applied to qubits. Amplitude damping is a collapse of a qubit into its lowest energy state, which causes a loss of its state and any entanglement with other qubits. This noise model is relevant to any qubit technology which stores its state in the energy levels of a quantum system, such as superconductors or trapped ions. The noise rate for amplitude damping is determined by the probability of collapse into the ground state [35]. Phase damping represents a loss of a qubit's phase. The combination of both is called amplitude and phase damping (APD). As qubits do not idle perfectly, and noise events will occur even if no logical gate was performed [36], we introduce Pauli and APD noise on each qubit on every cycle. Coherent noise, which comes from either external fields or improper calibration [5], can be particularly destructive [36].

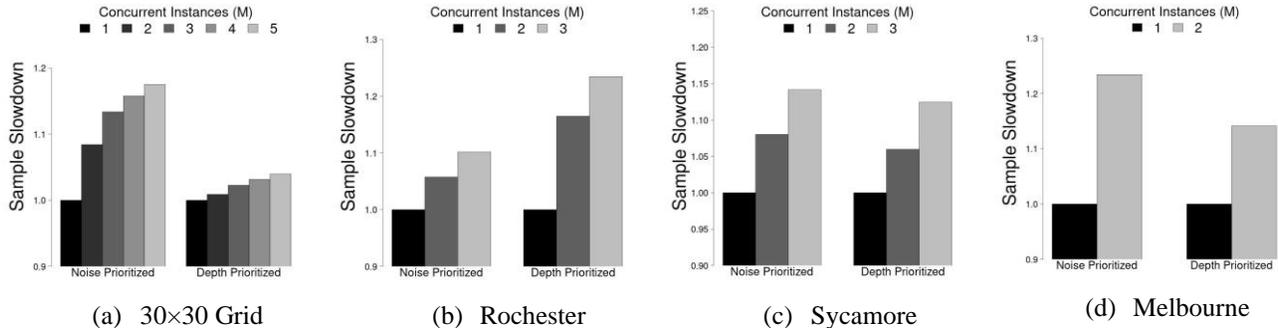

Figure 8: Effective sampling latency increase due to increase in number of instances (*M*) on different architectures. Results are shown for both noise prioritized and depth prioritized scheduling. *M* goes up to five on 30x30 Grid, up to three on Rochester and Sycamore, and up to two on Melbourne.

We model coherent noise as an over-rotation of each quantum gate, which emulates improper calibration. Along with each of the three quantum noise models we simulate measurement error. A measurement error corresponds to a bit flip on the final measurement. For example, a qubit may have collapsed to its $|1\rangle$ state, but a classical "0" was recorded.

Prior to simulation, the noise rates and measurement error rates for qubits on the QPU are generated uniformly at random, within a specified range. We based the Pauli and APD noise and measurement error rates on the Google Sycamore profiling data [2], where we matched the mean and standard deviations. The noise rates for single and two qubit gates are different, with noise for 2-qubit gates being significantly higher. All noise parameters are listed in Table 1. We use the same noise parameters for all QPU architectures.

### 5.1.1. Noise from concurrency

It was previously noted that running multiple instances will lower the average qubit reliability, as utilizing more qubits inherently requires using some of the lower quality qubits. However, running multiple instances simultaneously will degrade quality further by introducing additional noise, such as increased crosstalk. To capture this effect, we modify our noise rates to account for concurrent execution. We again base this on experimental data from Google's 52-qubit Sycamore processor [2]. Experimental profiling, based on cross-entropy benchmarking [4], provided the noise rates for operations performed in isolation and for the same operations when performed simultaneously. Relative to isolated, the single qubit gate error increased by 6.67%, the 2-qubit gate error by 43.1%, and the measurement error by 23.6% when performed simultaneously.

To match these results, we increase our base noise rates (in Table 1) by the respective amounts. When the QPU is at 0% utilization, the noise rate is the base rate. When the utilization is 100%, each of the noise rates increase by the amount shown in [2]. For utilization between 0% and 100%, we linearly scale the noise between the base rate and the maximal amount. Note that 100% utilization does not mean a gate is performed on every qubit on every cycle, but that 100% of the qubits are dedicated to a currently running program.

### 5.2. Preprocessing and Qubit Mapping

Prior to the execution of the quantum application, the noise rates of qubits on the QPU must be profiled to perform effective mapping. While we use error rates based on Google's Sycamore, error rates for qubits are also supplied by IBM after calibration cycles, or these can be found by performing diagnostics just prior running [39]. Such error rates are high-level estimates of real noise processes. To enable this noise information to be used effectively, we run the Floyd-Warshall (FW) path-finding algorithm, which is slightly modified to account for the noise rates. This produces a distance matrix (DM), which provides the pair-wise distance between each qubit, and a reliability matrix (RM), which provides the pairwise reliability of paths between all qubits. The DM is constant for a QPU architecture but the RM changes depending on the noise profile. For two qubits, qubit $i$ and qubit $j$, $DM[i, j]$ is the minimum distance between them and $RM[i, j]$ is the reliability (fidelity) of the most reliable path (chain of 2-qubit gates) between them. $m[j]$ is the measurement reliability of qubit $j$.

To run the circuit instances on the QPU, physical qubits need to be allocated to each one (i.e., each circuit instance must be mapped onto the QPU). The host processor uses a graph search algorithm and the pre-computed DM and RM to find the estimated best qubit allocation. It searches among the available qubits on the QPU for a subset of qubits (subgraph) with high reliability and connectivity. As we are mapping multiple instances of a single program (for a single user), we do not have to worry about maintaining fairness between instances, as is required for previous multiprogramming strategies [10][23]. Additionally, our mapping strategy is highly parallelizable and will remain efficient even for very large QPU sizes.

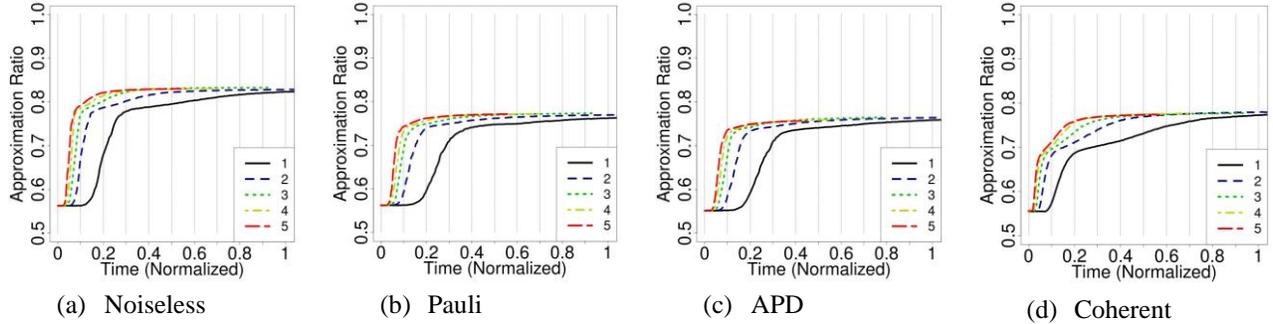

(a) Noiseless  (b) Pauli  (c) APD  (d) Coherent

**Figure 9: Training of 12-qubit QAOA on 30×30 Grid architecture with different noise models.**

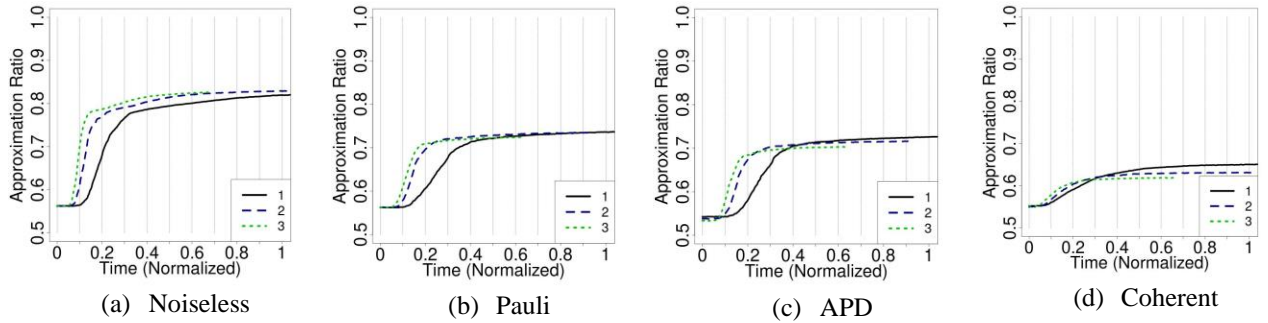

(a) Noiseless  (b) Pauli  (c) APD  (d) Coherent

**Figure 10: Training of 12-qubit QAOA on Rochester architecture with different noise models. Only three instances of the 12-qubit problem could be consistently mapped onto the 53-qubit architecture.**

The search is a variant of breadth-first search (BFS) which is greedy, and heuristic based. A single search starts on an available qubit, which is a subgraph (S) of size one. It then then expands S onto its available neighboring (candidate) qubits (C) until it is sufficiently large to run the quantum circuit. While searching, the algorithm uses three heuristics to choose which qubits to add to the subgraph:

1. Distance of all paths between the new qubit and every other qubit already in the subgraph.
2. Reliability of all paths (2-qubit gates) between the new qubit and every other qubit already in the subgraph.
3. Measurement error rate of the new qubit.

Many such subgraph searches are performed in parallel (one for each available starting point), and the subgraph which achieves the highest score on the heuristics is chosen. The pre-computed DM and RM enable this to be done efficiently. We use two variants of this search. The first prioritizes the reduction of noise rate, which considers heuristic 2 to be the most important. It adds qubits which have the overall most reliable connections back to the subgraph, as in Equation 1. Heuristic 1 is considered indirectly, as the distance between qubits impacts reliability as well.

**Equation 1:** $argmax_{j \in C} \sum_{i \in S} RM[i,j] \times m[j]$

The second search variant prioritizes the minimization of circuit depth, which considers heuristic 2 to be the most important. It adds qubits which have the minimum distance paths back to the subgraph, as in Equation 2. If multiple candidate qubits have the same distance, the reliability in Equation 1 is used as the tie breaker. Considering distance to be more important than reliability maximizes connectivity, which reduces the number of swaps required–which reduces circuit depth. Our experiments found connectivity is generally more important than reliability. Both variants consider measurement error rates equally.

**Equation 2:** $argmax_{j \in C} \sum_{i \in S} \frac{1}{DM[i,j]} \times m[j]$

As each instance requires a qubit allocation, this search is repeated once for each instance. Due to the graph search being greedy, there is a degradation in the quality of each consecutive allocation. The noise rates of the physical qubits on the QPU will determine how many instances can be run simultaneously. The number of instances can be specified by the user, or the host can make an educated guess based on the size of the problem and the number of qubits available.

### 5.3. Qubit Routing via SWAP Insertion

Once the qubit allocations are complete, the QAOA circuit needs to be transformed to run on its given set of physical qubits. This includes introducing additional SWAP gates to enable communication between logical qubits that interact. We use the Sabre [21] algorithm, which uses heuristics to find an approximate solution. As each instance of QAOA can have a different topology (connections between its physical qubits), each circuit instance can have different lengths after the introduction of swaps.

## 6. Results and Analysis

### 6.1. Speedup

A summary of the speedup achieved for QAOA training on different architectures, with different noise models and qubit allocation schemes is shown in **Error! Reference source not found.**. Significant speedups are achievable on every architecture, even in the presence of noise. This indicates that increasing $M$ not only overcomes the increase in sampling latency (from to $T$ to $T'$) but also any noise induced slowdown of optimization (from $K$ iterations to $K'$). However, the presence of noise does tend to reduce the benefit of concurrency slightly, which will be further detailed in Section 6.3. On average, depth prioritized mapping produces a speedup 11.5% greater than noise prioritized mapping. This is intuitive, as VQAs are already noise tolerant. Prioritizing the creation of short depth circuit reduces latency of training iterations, improving speedup. Additionally, reducing the depth can also help reduce the impact of noise by reducing required time that the qubits must remain coherent.

Note that, due to qubit count limitations, we report data only up to three simultaneous instances on the Sycamore and Rochester architectures, which are running a 12-qubit QAOA. While it is possible to fit four instances onto each machine, which both have 53 physical qubits, the qubit mapping process does not always allow for this. Heuristics, which prioritize the minimization of noise and circuit depth, are used to map instances onto physical qubits. The remaining available physical qubits are not guaranteed to be suitable to run more instances, as they may exist in disconnected regions of the machine. Hence, only three instances were able to be reliably mapped simultaneously. For the same reason, only two simultaneous instances of the 4-qubit QAOA were used on the 14-qubit Melbourne architecture.

As the 30×30 Grid architecture has room for many more instances, we greatly increased the concurrency to 20 and 45 simultaneous instances. The results are shown in Figure 6. The results show that large numbers of instances can continue to increase the speedup, if enough physical qubits are available.

### 6.2. Sample Latency

As discussed in Section 4.5, increasing the number of instances ($M$) can increase effective sampling latency due to potential increases in the circuit depth. The relative sampling latency of the quantum circuit (sample times) are shown in Figure 8. The latencies in Figure 8 indicate a modest increase in the sample time with increasing numbers of instances.

The qubit allocation scheme has a noticeable impact on the sampling latency overhead. Switching from noise prioritized allocation to depth prioritized significantly reduces the overhead on the 30×30 Grid architecture. This is because the 30×30 Grid has more than enough physical qubits with high connectivity to service all scheduled instances. Choosing a highly connected set of physical qubits for each instance ensures low swap overhead and low depth. The inverse is true on the Rochester architecture, because it has only a few regions of high connectivity. The qubit allocation is greedy, and hence gives the highly connected regions to the first instance. The second and third instances mapped onto Rochester only have qubits with low connectivity available. Hence, there is a larger increase in depth (and hence the sampling latency) with increasing numbers of instances.

### 6.3. Training

Due to space limitations, we only show the full training results for depth prioritized qubit mapping scheme, which has superior performance to the noise prioritized mapping scheme. The training process of QAOA involves finding parameters for the quantum circuit which produce the best results. Hence, there is an increase of the quality of the output of the quantum circuit (the approximation ratio) over time. The approximation ratio plotted against time is shown in Figure 9 for the 30×30 Grid, Figure 10 for Rochester, Figure 11 for Sycamore. For each architecture, we show the training process with different noise models and a noiseless case. The approximation ratio starts low in all cases, as the QAOA circuits are effectively producing random guesses. After each training iteration, more ideal parameter values are found, which produce higher approximation ratios. After some time, the approximation ratio plateaus, and further improvements are not possible.

In all cases, the presence of noise reduces the maximum approximation ratio that can be achieved. Notably, noise shows a more significant impact on the Rochester and Sycamore architectures than on the Grid architecture. This is due to two reasons. The first is that the qubit utilization is higher on Rochester and Sycamore, the instances are more densely packed, leading to increased noise rates. The second is due to the available connectivity of qubits on each architecture. The Grid architecture has more than enough qubits to find low-depth, optimal mappings for each instance. Rochester and Sycamore tend to run out of qubits, and higher numbers of instances tends to significantly increase the circuit depth. This makes the circuits more susceptible to noise, which prevents them from achieving the same quality of output. This effect is most pronounced with coherent noise, which has a very significant impact on the Rochester architecture, which has lower connectivity than Sycamore.

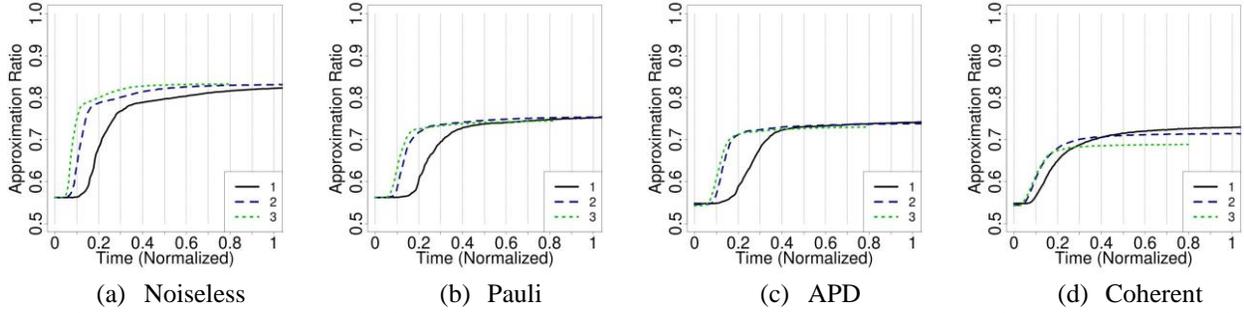

(a) Noiseless     (b) Pauli     (c) APD     (d) Coherent

**Figure 11:** Training of the 12-qubit QAOA on Sycamore architecture with different noise models. Only three instances of the 12-qubit problem could be consistently mapped onto the 53-qubit architecture.

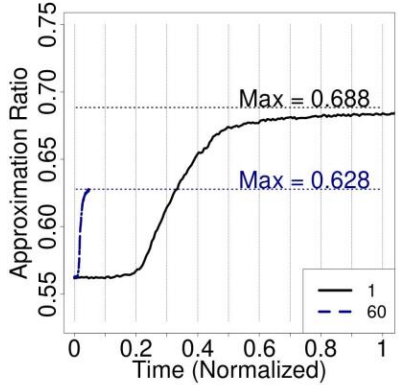

**Figure 12:** 30×30 Grid with Pauli noise increased by 10× for 2-qubit gates. Tested with $M = 1$ and $M = 60$, using 80% of the available qubits.

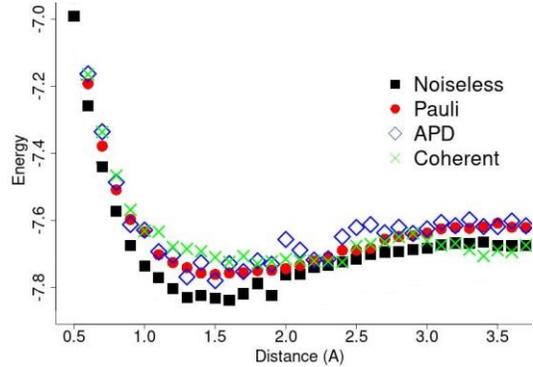

**Figure 13:** Minimum energy found by VQE for different distances between the Li and H atoms. Shown with noiseless simulation and with the noise models from Table 1. $M = 25$ used in all cases.

Coherent noise has a more significant impact with higher numbers of concurrent instances. This is because coherent noise due to mis-calibration of physical qubits will have a more consistent impact on each of the instances. This causes the distributions produced by each instance to be more different, making them more difficult to combine. Additionally, coherent noise is known to be particularly destructive [36].

### 6.4. Sensitivity to Variability

To find the sensitivity to large and highly variable noise, we increased the range of the 2-qubit noise rate by 10×. Hence, 2-qubit noise rates were set in the range 0.07-5.97%. We ran the 12-qubit QAOA on the 30×30 Grid with Pauli noise, with $M = 1$ instances and $M = 60$ instances. For the case of $M = 60$, 80% of the physical qubits were used. Hence, the mapping scheme is not able to avoid most noisy qubits. The results are shown in Figure 12. $M = 60$ was able to achieve a speedup of 10.9 over the $M = 1$ case, despite significantly increased noise and a $\frac{T'}{T}$ ratio of 1.67. $M = 60$ achieved a maximum approximation ratio of 0.65, whereas the $M = 1$ case was able to achieve 0.677. These results show resilience to large variances in noise. In addition to providing a large speedup, using multiple instances can help increase stability by averaging erroneous measurements. The optimal number of instances depend on noise, problem size, and QPU size.

### 6.5. Noise Induced Barren Plateau

QAOA uses repeating layers of parametric gates that represent the problem and driver Hamiltonian. Typically, the number of layers is denoted by $p$. In theory, increasing $p$ should increases the solution quality. However, increasing the number of layers ($p$) degrades the fidelity of quantum circuits running on the NISQ hardware. With more layers, gate count and circuit depth increase significantly. Increased circuit noise deteriorates the quality of the solution. This is known as the Noise Induced Barren Plateau (NIBP) [37].

To understand if the CQCS makes NIBP worse, we simulate QAOA circuit running in isolated mode ($M = 1$) and concurrent mode ($M = 8$). We observe that approximation ratios of both cases $M = 1$ and $M = 8$ are closely matched. This observation is consistent across 50-plus combinations of input circuits and qubit mappings. For this simulation, we use IBM's hardware-specific noise model that uses the Krauss operator obtained via gate set tomography. This model captures the device level variations, crosstalk, and decoherence.

## 7. Concurrent Execution of VQE

CQCS can be used to accelerate any VQA, including the Variational Quantum Eigensolver (VQE). As with QAOA, the main limitation for performance of VQE is the number of samples required [38]. We use CQCS and an efficient VQE implementation [19] to perform ground state estimation with the LiH molecule. VQE requires different Pauli measurements of the circuit, requiring the circuit to be re-run for every set of compatible measurements [25]. For LiH, there are 25 such sets, and hence 25 individual circuits that need to be run [19]. This significantly increases the number of samples that must be taken. We used CQCS to run all 25 circuits in parallel ($M=25$) on the 30×30 Grid architecture, using the same noise models and noise rates in Table 1. The hardware-efficient VQE implementation [19] allows the circuit to be reconfigured to match any qubit connectivity, thus no additional swap gates were required. Hence, the use of multiple instances did not increase the sample latency, as discussed in Section 4.5. This allowed the full 25× speedup of the quantum sampling. The minimum energies found by VQE are shown in Figure 13, which is in good agreement with results reported in [19].

Optimization for VQE can take hours, a time scale at which considerable drift in the gate parameters can occur [19]. Hence, it may be necessary to re-calibrate during the experiment [19]. CQCS can help by reducing the latency, which will reduce the amount of drift that occurs between each iteration. Unfortunately, the high accuracy required by VQE will also make it more susceptible than QAOA to the increased noise from CQCS. A potential strategy is to utilize CQCS initially to achieve speedup, but then to reduce parallelism once near the optimal solution to maximize accuracy.

## 8. CQCS on Neutral Atoms

CQCS can provide additional benefits on emerging Neutral Atoms (NA) platforms that are expected to scale to hundreds of qubits. In the ISCA 2021 paper, Baker et al. discussed 11 the potential opportunities and challenges in building runtime system for Neutral Atom (NA) architectures [3]. There are two technology-specific challenges in training QAOA or VQE on NA platform: **(1)** NA gates are significantly (10× to 100×) slower compared to superconducting qubits **(2)** NA suffers from Atom loss problem. We believe CQCS can improve the performance and reliability of NA systems by leveraging concurrent circuit sampling. CQCS can generate more samples per unit time to enable faster training on neutral atom platforms. Moreover, NA-based architectures have substantially less variability in the gate errors, enabling packing more copies per machine, enabling significantly shorter runtimes. For Neutral Atom systems, the use of CQCS can be significantly better than multiprogramming.

**Fighting Atom loss with concurrency:** Atom loss is a significant hurdle in scaling NA systems. As neutral atom qubits can escape from optical tweezers and drift into the environment, the total number of physical qubits on the machine can reduce due to atom loss. There is about 1% chance that the physical qubit might be lost upon measurement. This can be solved by reloading the physical qubits [3][20]. Unfortunately, reloading a qubit is slow, it can take hundreds of milliseconds to reload the qubit. A frequent reloading of qubits can significantly slow down the execution. CQCS can help tolerate atom loss errors and delays. As we use multiple copies for sampling, atom loss error on one copy can be tolerated as the rest of the copies can still produce the output samples, and thus, execution can continue without stalling.

## 9. Related Work

Multiprogramming for quantum computers was introduced in [2]. The authors suggest heuristics for trading off gate and measurement errors for program allocation. In addition, they develop methods to detect when MP is causing degradation in the results. Further multiprogramming mapping strategies were developed in [23]. Additionally, the authors of [23] develop inter-program SWAP techniques which can reduce the number of swaps required. The provides an advantage, at the risk of increased cross talk between co-running programs. Methods for accelerating simulation of VQAs in the presence of noise have been developed in [17]. Such strategies could be used to accelerate the simulations used in this work. Parameters of QAOA are commonly initialized randomly. However, it is possible to use predictive strategies to find near optimal initial parameters for some problems [40], reducing the number of training iterations required.

## 10. Conclusion

We have shown that concurrent quantum circuit sampling (CQCS) can speedup for modern VQAs, such as QAOA, and VQE while allowing for nearly full QPU utilization. These benefits can be achieved even in the face of various realistic noise models.